\documentclass[aps,prd,superscriptaddress]{revtex4-1}
\usepackage{graphicx}
\usepackage{xspace}
\usepackage[usenames,dvipsnames]{color}  

\newenvironment{eqn}[1]{  \newcommand{\passarg}{#1}  \begin{equation}
    \ifshowkeys\key{#1} \ \ \fi}{\label{\passarg}  \end{equation}}
\newcommand{\beq}[1][]{\begin{eqn}{#1}}
\newcommand{\eeq}{\end{eqn}}
\newcommand{\degg}{\ensuremath{\mathrm{deg}^2}\xspace}
\newcommand{\chieff}{\ensuremath{\chi_\mathrm{eff}}\xspace}
\newcommand{\eq}[1]{Equation~\ref{#1}\ifshowkeys\key{(#1)}\fi}

\newenvironment{cfig}[1]{  \newcommand{\passarg}{#1}  \begin{figure}[htb] \centering
    \ifshowkeys \key{#1} \\ \fi}{\label{\passarg} \end{figure}}
\newcommand{\bfig}[1][]{\begin{cfig}{#1}}
\newcommand{\efig}{\end{cfig}}
\newcommand{\fig}[1]{Figure~\ref{#1}\ifshowkeys\key{(#1)}\fi}

\newcommand{\img}[2][0.75]{ \includegraphics[width=#1\textwidth]{#2}}

\newcommand{\sect}[1]{Section~\ref{#1}\ifshowkeys\key{(#1)}\fi}
\newcommand{\seclabel}[1]{\label{#1}\ifshowkeys\key{#1} \\ \fi }

\newcommand{\Jnote}[1]{{\color{green} \ifshownotes #1\fi}}
\newcommand{\key}[1]{{\color{blue} \rm{#1}}}
\newcommand{\citekey}[1]{\cite{#1}}

\newcommand{\vtheta}{\ensuremath{\vec{\theta}}\xspace}

\newif\ifshowkeys
\showkeysfalse

\newif\ifshownotes
\shownotesfalse  

\begin{document}

\title{Improving astrophysical parameter estimation via
offline noise subtraction for Advanced LIGO}

\author{J.~C.~Driggers}
\affiliation{LIGO Hanford Observatory, Richland, WA 99352, USA}

\author{S.~Vitale}
\affiliation{LIGO, Massachusetts Institute of Technology, Cambridge, MA 02139, USA}

\author{A.~P.~Lundgren}
\affiliation{Max Planck Institute for Gravitational Physics (Albert
  Einstein Institute), D-30167 Hannover, Germany} 

\author{M.~Evans}
\affiliation{LIGO, Massachusetts Institute of Technology, Cambridge,
  MA 02139, USA}

\author{K.~Kawabe}
\affiliation{LIGO Hanford Observatory, Richland, WA 99352, USA}

\author{S.~E.~Dwyer}
\affiliation{LIGO Hanford Observatory, Richland, WA 99352, USA}

\author{K.~Izumi}
\affiliation{LIGO Hanford Observatory, Richland, WA 99352, USA}

\author{R.~M.~S.~Schofield}
\affiliation{University of Oregon, Eugene, OR 97403, USA}

\author{A.~Effler}
\affiliation{LIGO Livingston Observatory, Livingston, LA 70754, USA}

\author{D.~Sigg}
\affiliation{LIGO Hanford Observatory, Richland, WA 99352, USA}

% ...

\collaboration{The LIGO Scientific Collaboration Instrument Science Authors}

\begin{abstract}
The Advanced LIGO detectors have recently completed their second observation run successfully. The run lasted for approximately 10~months and lead to multiple new discoveries. The sensitivity to gravitational waves was partially limited by correlated noise. Here, we utilize auxiliary sensors that witness these correlated noise sources, and use them for noise subtraction in the time domain  data. This noise and line removal is particularly significant for the LIGO Hanford Observatory, where the improvement in sensitivity is greater than 20\%. Consequently, we were also able to improve the astrophysical estimation for the location, masses, spins and orbital parameters of the gravitational wave progenitors. 
\end{abstract}


\maketitle

\section{Introduction}
\seclabel{sec:Intro}

Advanced LIGO's~\citekey{aLIGOdescription} detections of gravitational waves~\citekey{GW150914_paper, GW170817_BNS_paper, MME, GW151226-DETECTION, GW170104_paper, GW170814_paper, GW170608_paper} have opened up a new view of the universe, allowing us to learn about astrophysical sources such as the mergers of compact stellar remnants.  As work continues toward reaching the design sensitivity of Advanced LIGO, we are looking forward to more detections of gravitational waves (GW), and learning more about their sources~\citekey{ObservScenarios, ObservScenariosPublished}. 

The Advanced LIGO detectors are kilometer-scale laser interferometers with suspended test masses, sophisticated seismic isolation systems and complex optical configurations employing multiple coupled optical resonators~\citekey{TheLIGOScientific:2016agk}. 
LIGO's design sensitivity is limited primarily by fundamental noise sources such as quantum shot noise, quantum radiation pressure noise, and Brownian thermal noise.  However, in some frequency bands, Advanced LIGO's first observation runs have been limited by technical noises~\citekey{O1-NOISEPAPER, GW170104_supplement}. Many of these noise sources are well understood, and further work is needed to prevent them from contaminating the gravitational wave sensitivity. In practice, a balance has to be struck between commissioning the detector to improve noise performance, and observations. For the second observation run from November~2016 to August~2017, the LIGO Scientific Collaboration elected to run with somewhat elevated noise in one of the interferometers, while making plans to address these noise sources prior to the third observation run. 



The sensitivity of the LIGO Hanford detector is severely affected by laser noise in the frequency band from 100~Hz to 1~kHz. Fortunately, we have a set of independent witness sensors that is highly correlated with this noise. The spectra of both detectors also reveal the power mains and its harmonics, as well as lines that are monitoring the calibration in realtime.
Noise and line removal can enhance the sensitivity of the LIGO Hanford detector by more than 20\,\% during the second observing run. By implementing a post-processing noise removal algorithm, we are able to significantly improve our ability to estimate the parameters of a compact binary coalescence, including its sky location, distance, masses, spins and orbital mechanics.


The removal of lines with narrow frequency spread is also beneficial in the search for continuous wave sources, such as spinning neutron stars. Due to the long signal duration, the frequency of the observed gravitational wave signals are Doppler shifted by the rotation of the earth and the motion of the earth around the sun. Narrow spectral lines are broadened after the Doppler effect has been removed, and can therefore impact a wider frequency band.
Currently, the continuous wave searches ignore a relatively wide frequency gap around each spectral line to account for this. The search for continuous waves is fundamentally a post processing algorithm and can naturally benefit from any sensitivity improvements made in our post-processing noise and line removal technique.

\sect{sec:noisesources} briefly describes the origin of several noise sources that can be removed, while \sect{sec:method} discusses the available witness sensors and the method for calculating the coupling functions used in the subtraction algorithm.
\sect{sec:PE} examines how this post-processing noise subtraction impacts the estimation of various astrophysical parameters.

\section{Technical noise sources}
\seclabel{sec:noisesources}

After the first Advanced LIGO observing run, lasting from September~2015 to January~2016, the two LIGO observatories in Hanford, Washington, and Livingston, Louisiana, underwent a series of upgrades~\citekey{GW170104_supplement}. The Handford observatory focused on increasing the amount of laser power circulating in the interferometer, which required using a high power oscillator~\citekey{HPO_doc}. The water required for cooling the laser rods flows through piping attached to the laser table, causing vibrations. This table also hosts an optical train for shaping the beam, impressing radio frequency sidebands, beam steering, and frequency stabilization. The water flow causes vibrations of these optical elements, which translates into beam jitter. Thermal fluctuations in the laser rods and mode mismatch in the optical resonators are also causing jitter variations of the the laser beam size---likely due to the turbulent water flow directly over the laser rods~\citekey{JitterAlog}.

During Advanced LIGO's second observing run, it was discovered that one of the core mirrors in the long arm cavities at the Hanford observatory sports a point absorber on its surface. Depending on the incident power, this causes a beam deformation due to the thermal expansion of the optics and the temperature dependence of the index of refraction. LIGO has a thermally actuated adaptive optics system~\citekey{TCS_doc} to help compensate for effects that are axially symmetric about the beam axis, but which cannot remove the effect of a point absorber that is located several millimeter away from the center. 

The presence of this non-asymmetric optical deformation couples beam jitter and beam size variations into the gravitational wave readout channel. During the run, an attempt was made to inspect the offending mirror and try to clean off the absorption spot. But, this failed with the cleaning having no appreciable effect on the absorption, and so this optic has been replaced prior to the start of the third observing run. As a result, the higher beam jitter and beam size variations significantly impacted the gravitational wave readout sensitivity of the LIGO Hanford Observatory during the entire second observation run.

The LIGO Livingston Observatory did not utilize their high power laser oscillator during the second observation run, and so required much less cooling water to flow through the piping on the laser table. Instead, the commissioning effort at the Livingston Observatory focused on finding and mitigating various noise sources, such as scattered light and coupling of electronics noise. This complementary approach in the commissioning of the two LIGO detectors is common, and enables early experience with new hardware configurations. Both interferometers are susceptible to the power mains that show up in the gravitational wave readout. An active set of calibration lines is used to track the variations in the optical gain. Both the power mains and the calibration line are well known, and therefore can be subtracted from the measured gravitational wave strain.

At frequencies below a few tens of Hertz, additional noise is introduced by the control forces that are applied to the mirrors to control the resonance condition of the interferometer and the test mass orientation. In addition to the 4\,km long arm cavities, the LIGO detectors have optical cavities in the central part of the interferometer whose lengths must be controlled to keep the interferometer at its linear operating point. The sensors used for the auxiliary length degrees of freedom have worse shot noise limited sensitivity than the gravitational wave readout channel. Since this shot noise is imposed on the actual length noise of the auxiliary cavities by our feedback controls, it can contaminate the gravitational wave readout channel. The feedback control system actively tries to decouple two of the three length degrees of freedom from the gravitational wave readout channel. But, due to a remaining imbalance in the actuator strength and due to leaving the third degree of freedom untouched, some noise still couples into the gravitational wave readout channel. This noise can also be removed in post-processing.


\fig{fig:Noises} shows the noise amplitude spectral density (ASD) of the LIGO detectors in the low-latency readout, and estimates the noise contributions from each of the categories described above.

\bfig[fig:Noises]
\img[1]{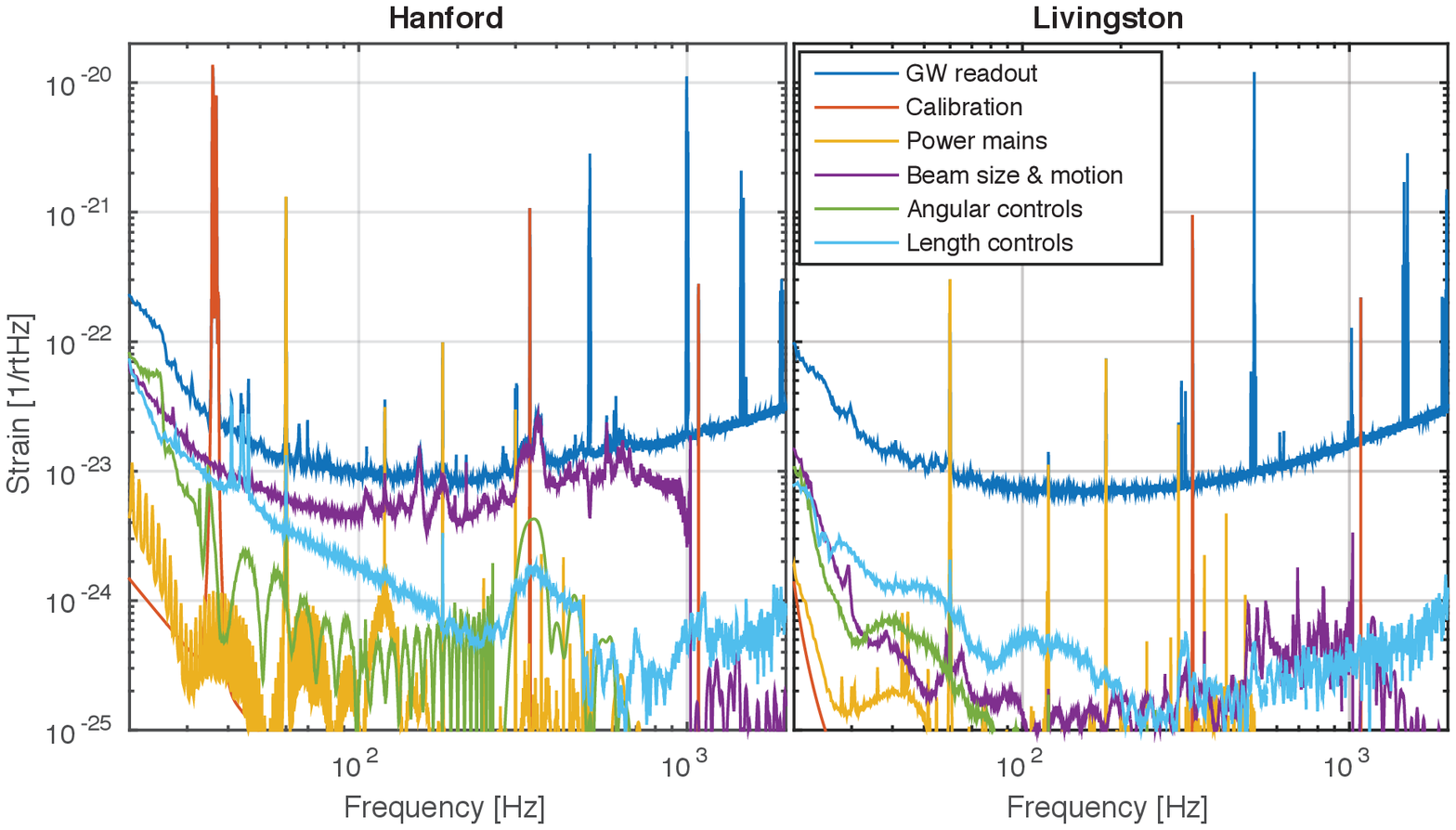}
\caption{Noise amplitude spectral density of the Advanced LIGO detectors (dark blue) with the left panel for Hanford and the right panel for Livingston. The other traces are the estimated contributions of the calibration lines (red), power line and harmonics (gold), beam jitter motion and beam size variations (purple), angular control noise (green), and the auxiliary length controls degree of freedom (light blue). These spectra are based on 1024\,s of data starting on 25 June 2017 at 08:00:00\,UTC, at a time when both LIGO interferometers were operating and in an observation ready state.}
\efig

\section{Noise subtraction}
\seclabel{sec:method}


We use the optimal Wiener method ~\citekey{norbert} to estimate the coupling function
between each noise source and the GW channel.  This method
determines how best to manipulate an auxiliary witness sensor's data
such that when it is subtracted from the primary target signal (here, the GW
channel) the mean-square-error of the primary channel is minimized.
To do this, we define an error signal
\beq
\vec{e} = \vec{d} - \vec{y},
\eeq
where $\vec{d}$ is the noisy target signal and
$\vec{y}$ is the approximation of $\vec{d}$ from the independent witness
sensor.  This is given by
\beq
\vec{y} = \vec{w}^{\rm T} \vec{x},
\eeq
where $\vec{x}$ is the measurement of the external disturbance from the
witness sensor, and $\vec{w}$ is the finite impulse response (FIR) filter
that we will solve for.
The figure of merit $(\xi)$
that we use for calculating the Wiener filter coefficients in
this case is the expectation
value of the square of the error signal,
\beq[eq:FOM]
\xi \equiv E[\vec{e}\,^2] = E[\vec{d}\ ^2] - 2 \vec{w}^\textrm{T} \vec{p} +
\vec{w}^\textrm{T} R \vec{w}.
\eeq

Here, $E[*]$ indicates the expectation value of $*$, $\vec{p}$ is the
cross-correlation
vector between the witness and target signals, and $R$ is the autocorrelation
matrix for the witness channels.
When we find the extrema of \eq{eq:FOM} by setting
\beq[eq:optimize]
 \frac{d \xi}{d w_i} = 0,
\eeq
we find
\beq[eq:Rwp]
 R \vec{w}_{\rm optimum} = \vec{p}.
\eeq

\eq{eq:Rwp} finds the time domain filter
coefficients which minimize the RMS
of the error $\vec{e}$ by optimizing the estimate of the transfer function
between the witness sensors and the target signal.
The error signal is now an estimate of the signal in $\vec{d}$, without any noise.

This method was
utilized on LIGO data in 2010, for low frequency seismic
noise~\citekey{iLIGO_SeisCleaning}.
Following this, the method was used to create feed forward
filters which were used online in 2010~\citekey{FFWpaper}.  This and
another method were also
used offline to remove noise from auxiliary degrees of freedom from
LIGO's initial-era sixth science run~\citekey{iLIGO_LSCCleaning,
  iLIGO_PEMCleaning}.  A frequency-dependent variant of noise subtraction was proposed in 1999~\citekey{FreqDepCleaning}, and shown to be effective on a prototype interferometer's data.  

The Wiener method is able to handle several witness sensors
simultaneously by extending $R$ and $\vec{p}$ in the above equations,
even if they see some amount of signal from the same
noise source, as long as the information in the auxiliary sensors is
not identical.  This
prevents over-subtracting a source of noise, and eliminates the need
to carefully chose the order of subtraction if the witness sensors are
used in series.
The inversion of the matrix
($R^{-1}$ when \eq{eq:Rwp} is solved for $\vec{w}_{\rm optimum}$) is
computationally intensive, and is the main time-limiting step in the
noise removal process.

As this method works to minimize the root mean square (RMS) of
the target channel, it is
useful to remove narrow spectral lines from the data before attempting to
subtract the broadband noise sources.
For the calibration lines we use the digital signals that
are sent to the various actuators as our auxiliary channels.  Since we
know that these signals sent to different actuators are not correlated
with one another, we subtract them in series.  For the
power mains line at 60\,Hz we use a digitized signal that comes
directly from monitoring the voltage supplied to our analog
electronics racks.  While we monitor the voltage at all locations that
host analog electronics for the interferometer, we empirically chose
the one signal at each site that removes most of the 60\,Hz line.  In
the future, we may consider utilizing more of these signals,
particularly for subtraction over longer periods of time.

To measure the beam jitter motion we use a set of three split
photodiodes, each with four sections.
One of the photodiodes is placed on the laser table, and monitors the
beam motion and beam size just after the laser itself.  This diode has
a central circle, and a ring of three equal-sized segments surrounding the central
region.
The other two split photodetectors monitor the
vertical and horizontal motion of the beam rejected by the input mode
cleaner cavity which  spatially filters the laser beam before it enters
the main interferometer.
The signals from these photodiodes are all passed to the Wiener filter
calculation algorithm together.  


For both the angular and length control
noise sources we use the digital control signals that are sent to the
mirror actuators as the witnesses.

\fig{fig:Spectrum} shows the improvement that can be made in the LIGO interferometers'
noise ASD, as a function of frequency.  Note that the LIGO Hanford
detector is compromised by the technical noise sources
discussed above more significantly than is the LIGO Livingston
detector, and so sees much more dramatic improvement.
Notable spectral lines such as those at $\sim$500\,Hz and harmonics cannot
be independently witnessed with currently existing hardware, and so
cannot be subtracted.
The LIGO interferometers' alignment can change slowly with time, which causes the coupling of some noise sources to change.  Empirically, it seems that the coupling functions should be recalculated about once per hour of data.  For all of the gravitational wave events that have utilized this method of post-processing noise subtraction (GW170608~\citekey{GW170608_paper}, GW170814~\citekey{GW170814_paper}, and GW170817~\citekey{GW170817_BNS_paper}), the coupling functions were calculated using 1024\,s of data, and applied to 4096\,s of data surrounding each event.   

A measure of the improvement in each interferometer can be summarized
by the increase in horizon distance the detectors can see a certain GW
signal with a pre-defined signal to noise ratio.  For both canonical
binary black hole $30\,M_\odot$-$30\,M_\odot$ mergers as well as
canonical neutron star $1.4\,M_\odot$-$1.4\,M_\odot$
coalescences with a signal to noise ratio of 8
the Hanford detector improves by more than 20\% while the
Livingston interferometer only improves by about 0.5\% using this
measure.

\bfig[fig:Spectrum]
\img[1]{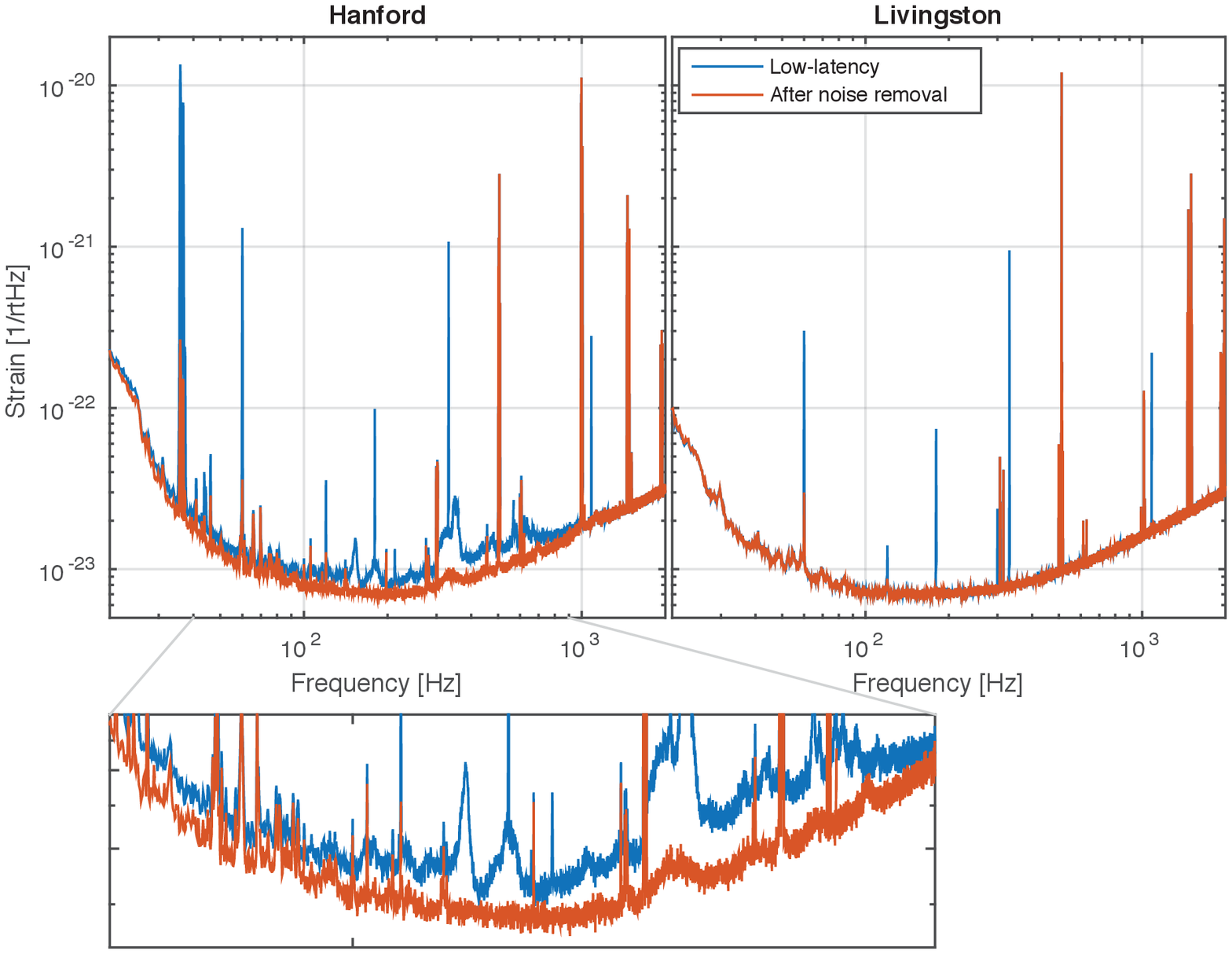}
\caption{Noise amplitude spectral density improvement of the LIGO detectors, Hanford in
  left panel and Livingston in right panel.  Low latency data used to
  identify GW candidates and determine their significance shown in
  blue traces.  Interferometer noise ASD after post-porcessing noise
  removal shown in red traces. Inset is zoom of Hanford data.
These
  spectra are estimated using 1024\,s of data starting on 25 June 2017
at 08:00:00\,UTC, a time when both LIGO interferometers were online
and in an observation ready state.}
\efig

Several checks can be done to confirm that this noise removal
procedure does not affect any gravitational wave signal present in the
data.  Most of the noise sources have no possibility of
containing any gravitational wave information, therefore cannot remove
any actual signal.  For example, the power mains monitors, calibration
lines, and beam motion photodiodes do not contain any
GW signal.
For other witness sensors such as length of the short Michelson, we
can calculate that the GW signal there is a factor of
$1.2 \times 10^{-5}$ smaller than in the main GW readout
channel~\citekey{DriggersThesis, iLIGO_LSCCleaning,
  Izumi_IFOresponse}, so should
only impact the GW signal up to 0.0012\%.

Perhaps the most robust way to check that there is no harm done to a
GW signal is to examine software injected signals, where we know the
true (simulated) parameters, and can compare the estimated parameters
from the low-latency data and the post-processed noise-subtracted
data.
We do this in \sect{sec:PE}. 

\section{Estimation of astrophysical parameters}
\seclabel{sec:PE}

Improved interferometer sensitivity not only enhances our confidence
that a signal is of astrophysical origin, but it greatly improves our ability to
estimate astrophysical parameters associated with the source of the
GWs.  

To estimate quantitatively the impact of post-processing
noise subtraction on the characterization of GW sources, we performed software injections of signals emitted from compact binary coalescences.
By software injection one means a simulated GW signal which is added to either real or synthetic interferometric noise.

For this study, we created 10 binary black hole (BBH) signals and 9 binary neutron star (BNS) signals. 
For the BBH, we used the { \tt IMRPhenomPv2} waveform model,
 whereas for the BNS we used {\tt TaylorF2}.
 This latter does not include merger and ringdown, which is a reasonable approximation at low masses.
The BBH have component (detector-frame) masses randomly drawn from the range $[28- 64]\,M_\odot$, resulting in mass ratios
~\footnote{We define the mass ratio $q\equiv m_2/m_1$, where by convention $m_1\geq m_2$.} 
in the range $0.48-1.0$. The BBH are injected with zero spin (although we \emph{do} allow for the spins degrees of freedom while measuring the BBH parameters).
The BNS have masses in the range $[1.41 - 1.45]\,M_\odot$ and no spins.
The luminosity distance of the events random in comoving volume. In practice, this results in distances between 70\,Mpc and 1.54\,Gpc for the BBH and 14\,Mpc and 138\,Mpc for the BNS.

For each signal, we add it
to a stretch of the LIGO Hanford and Livingston detectors data when both instruments were online and in a nominal
observational state (1024\,s beginning at 25 June 2017 08:00:00 UTC). For each signal and each instrument, a frame file (this is the file format commonly used within the LIGO-Virgo Collaboration to store GW data
containing the signal and the original noise is created and saved.
The cleaning procedure is then performed on all frame files, and new ``cleaned'' frames  are stored.
This leaves us with two sets of frame files: one with the original LIGO data (as well as the GW signal), and one with the cleaned data (and, again, the signal).

Both set of frames are analyzed with the same algorithm used by the LIGO and Virgo collaborations to characterize compact binary coalescence sources,  { \tt LALInference}~\cite{2015PhRvD..91d2003V}.
For each signal, we aim to obtain a posterior distribution for the unknown parameters on which it depends, \vtheta, given the stretch of data containing the i-th signal: $p(\vtheta| d_i)$.
Using Bayes' theorem, this can be written:
\begin{equation}
p(\vtheta| d_i) \propto  p(d_i | \vtheta ) p(\vtheta),
\end{equation}
where the proportionality coefficient just acts as an overall normalization. The first term on the right hand side, $p(d_i | \vtheta )$, is the likelihood of the data given the parameters. In this paper we work with a two-detector network. Assuming noise is statistically independent in the two instruments we can write the network likelihood as the product of the likelihood in each instrument:
\begin{equation}
 p(d_i | \vtheta )=\prod_{k=1}^{\mathrm{N\;IFO}} p(d_i^k | \vtheta ).
\end{equation}
Finally, $p(\vtheta )$ is the prior distribution of \vtheta. For this study we used the same priors already utilized by the LIGO and Virgo collaborations, see e.g. Ref.~\cite{GW150914-PARAMESTIM}. 

Most of the unknown parameters are common to both the BBH and the BNS analyses. These include component masses, luminosity distance, orbital inclination and polarization, sky position, arrival time, and phase~\cite{GW150914-PARAMESTIM}. 
The only difference is that for the BNS analysis we assumed spins are aligned with the orbital angular momentum, while for the BBH we allowed for the possibility of misalignment, and hence precession. For the BBH runs, we relied on the reduced order quadrature approximation to the likelihood~\cite{Smith:2016qas} to reduce the runtime.

On average, we observe an increase of the signal-to-noise (SNR) at Hanford of $\sim 29\%$ for both BNS and BBH. This results on a $\sim 10\%$ increase in the \emph{network} SNR. The improvement in the network SNR is less dramatic, since the SNRs at each interferometer are added in quadrature~\cite{2015PhRvD..91d2003V}.

The increased SNR at Hanford yields a more balanced distribution of the SNR across the two sites, which mostly helps measuring the location of the source on the sky.
For the BNS simulations, we find a $32.2\%$ reduction of the 90\% credible interval in the sky localization, compared to what was obtained with the original uncleaned data. This is shown in Fig.~\ref{fig:SkyImprovement} for the loudest BNS we simulated
~\footnote{Most of the other events show similar improvements; we chose to show this one because for other sources the skymap has a characterizing ring-like structure, which does not allow one to clearly see the difference between the reconstructions.}
. The green curves refers to the analysis using the original data, whereas the blue curves are obtained with the cleaned data. 50\% and 90\% contours are given, and and star shows the true position. For this BNS, the 90\% sky area decreases from 39.6~\degg to 11.6~\degg using cleaned data, while the SNR in Hanford increases from 25.5 to 33.0 (the network SNR increases from 65.0 to 68.5).
For BBH, the average reduction of the 90\% sky uncertainty is $19.5\%$.
The estimation of the sources' luminosity distance is also improved, although not dramatically so, since its measurement requires detection of both GW's polarizations, whereas the two LIGO sites have nearly aligned arms. Adding more SNR in Hanford thus does not add significant polarization information. We find an average improvement of $6.6\%$ for BNS and $2.0\%$ for BBH for luminosity distance.

The the intrinsic parameters of the sources, i.e. masses and spins, can usually be measured well already with a single instrument. The uncertainty in those quantities is thus mostly affected by the network SNR, and not so sensitive to how that SNR is distributed in the network.
For the chirp mass~\cite{GW150914-PARAMESTIM} we find a relative improvement of $5.2\%$ for the BNS and $13.4\%$ for the BBH. The reason why BBH improve more is because they have fewer inspiral cycles (from which the chirp mass is measured~\cite{GW150914-PARAMESTIM}) than BNS. They can thus benefit more from any extra SNR at frequencies below $\sim 100$~Hz.

Similar small improvements are visible for the asymmetric mass ratio and the effective spin parameter \chieff~\cite{2008PhRvD..78d4021R,GW150914-PARAMESTIM}.
We find that cleaning makes little difference in this case. For a few BNS sources the uncertainty is actually slightly smaller before cleaning. On average, the mass ratio is estimated $1.4\%$ \emph{worse} when using cleaned data.
For BBH, we also find a few sources for which the uncertainty is smaller before cleaning, although when averaging over all sources, cleaned data yield intervals which are $5.1\%$ better with cleaned data. 
For \chieff we find an average improvement of $8.5\%$ for BBH and $0.1\%$ for BNS.

\bfig[fig:SkyImprovement]
\img[.95]{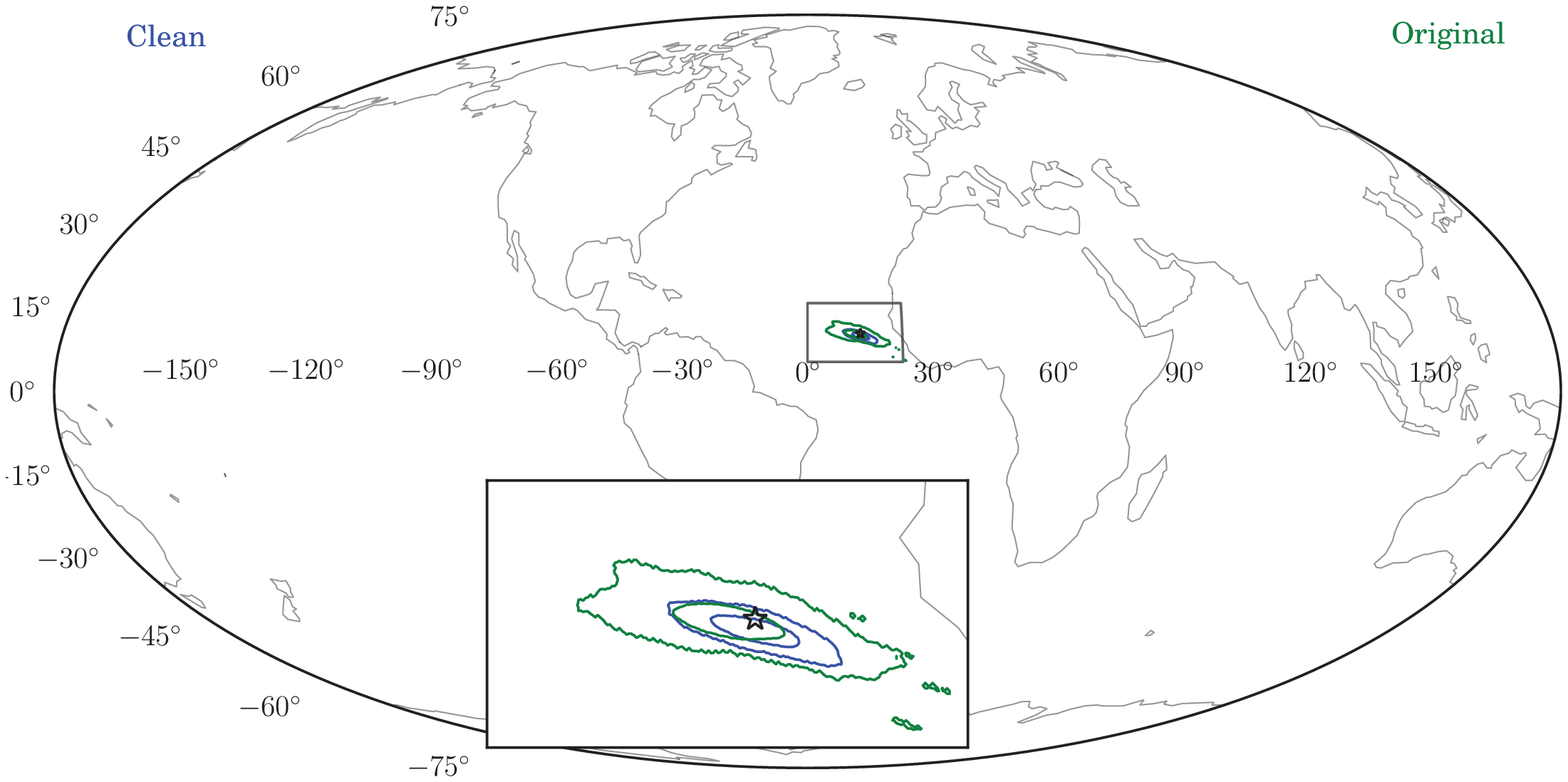}
\caption{Mollweide projection for the skymap of one of the BNS sources as measured with the original (green) and cleaned (blue) data. For each map the contours show the 50\% and 90\% confidence intervals. A star shows the true position of the injected source. To enhance clarity, the inset shows a zoom of the relevant part of the sky. The 90\% confidence interval decreases from 39.7~\degg to 11.6~\degg after data is cleaned.}
\efig

\section{Conclusions}
\seclabel{sec:conclusion}

We have shown that post-processing noise subtraction is effective for
the Advanced LIGO gravitational wave detectors, particularly for the
Hanford Observatory during the second observation run which is limited by known technical noise  sources
over a wide range of frequencies.
This sensitivity
improvement significantly enhances our ability to extract
astrophysical information from our detected signals.  
The improvement in sensitivity shown
roughly doubles the volume of the universe in which the Hanford
interferometer can detect gravitational waves.  Currently underway is the rewriting of the noise subtraction code such that it is scalable and can be applied to the entirety of the data from the second observation run.  This will allow the background estimations of events to also use cleaned data, and will enable the CW search to look at frequencies they have never been able to see before.

\Jnote{While we hope that future observation runs will be mostly limited by fundamental noise sources that are not correlated with externally-measureable noise sources, we plan to apply this noise removal technique to future observation runs.  This will be particularly useful if, as in the case of the laser jitter noise, we identify a technical noise source during an observing run and can install a witness for that noise, but cannot practicably remove the noise source on a short timescale.  The removal of mains lines and calibration lines will be implemented, which helps CW and makes PSD estimation easier, so other pipelines can do better.}

\section{Acknowledgements}
\seclabel{sec:ack}
The authors would like to thank the LIGO Scientific Collaboration's astrophysical parameter estimation group for their support.  
SV would like to thank R.~Essick for providing the code to plot the sky location of  sources.   
We are also very grateful for the computing support provided by The MathWorks, Inc.
LIGO was constructed by the California Institute of Technology
and Massachusetts Institute
of Technology with funding from the National Science
Foundation and operates under cooperative agreement
PHY-0757058.  This article has been given LIGO
document number P1700260.

\noindent\rule{4cm}{0.4pt}


%

\end{document}